\documentclass[11pt]{article}
\usepackage{moriond, epsfig}

\def\be{\begin{equation}}
\def\ee{\end{equation}}
\def\bea{\begin{eqnarray}}
\def\eea{\end{eqnarray}}

\begin{document}
\vspace*{4cm}
\title{Data-driven Estimation of SM Backgrounds for SUSY Searches at the LHC}

\author{Takayuki Yamazaki,\\
on behalf of ATLAS and CMS collaborations}

\address{Graduate School of Science, University of Tokyo\\
7-3-1 Hongo, Bunkyo-ku, Tokyo 113-0033, Japan}

\maketitle
\abstracts{Searches for supersymmetry (SUSY) are very important tasks at the Large Hadron Colleder 
(LHC). If SUSY exists at the TeV scale, clear excess above the Standard Model (SM) background will be 
observed. SM background should be estimated from real data self. 
In this paper, we descrive the strategy for the early SUSY searches at the LHC and focus on the data-driven 
estimation of the SM background in the early stage of collision.
}

\section{Introduction}
The Standard Model (SM) can explain almost all experimental results so far but is not considered to be a 
complete theory. Supersymmetry (SUSY), which is symmetry between fermions and bosons, is the most attractive 
theoretical model for new physics beyond the SM. 
If SUSY particles exist at the TeV scale, the hierarchy proplem can be resolved and three forces, 
electromagnetic, weak and strong interaction are unified at the GUT scale ($\sim10^{16}$GeV).
Moreover, the lightest SUSY particle (LSP) is a good candidate for the cold dark matter if R-parity is 
conserved.\\
\indent At the Large Hadron Collider (LHC), which will start to operate in 2008. 
Direct searches of the physics at the TeV scale can be performed. 
The strategy for early SUSY searches is described in this paper, and I focus on the 
estimation of SM background using the data corresponding an integrated luminosity of 1fb$^{-1}$.

\section{SUSY Signitures}
Sparticles such as gluinos ({\it \~{g}}) and squrks ({\it \~{q}}) will be produced via strong 
interaction at the LHC. They decay into the lightest state of the SUSY particle and SM particles. The lightest 
SUSY particle is stable if R-parity is conserved. Final event topology is high $p_T$ multijets and large mising 
$E_T$, sometimes including leptons. This is model-independent signiture of the SUSY events, and our SUSY search 
is based on this experimental signiture.\\
\indent Search modes are classified by the number of isolated leptons. The prommising search modes are 
listed in Table 1. They cover realistic SUSY models such as minimal supergravity (mSUGRA), gauge 
mediated SUSY breaking (GMSB), and anomaly mediated SUSY breaking (AMSB).

\begin{table}[h]
\caption{Search modes, corresponding SUSY modles and SM background processes.\label{tab:modes}}
\vspace{0.4cm}
\begin{center}
\begin{tabular}{ccc}
\hline \hline
number of leptons & SUSY model & SM background \\
\hline
no lepton  & mSUGRA, AMSB & QCD, $t\bar{t}$, $W$, $Z$ \\
one lepton ($e$, $\mu$) & mSUGRA, AMSB & $t\bar{t}$, $W$ \\
dilepton ($e$, $\mu$) & mSUGRA, AMSB, GMSB & $t\bar{t}$ \\
\hline \hline  
\end{tabular}
\end{center}
\end{table}

\indent The selection criteria of SUSY searches are

\begin{itemize}
\item 4 jets with $p_T >$ 50 GeV/c and at least 1 jet with $p_T >$ 100 GeV/c
\item $S_T >$ 0.2
\item $E^{miss}_T >$ 100 GeV and $E^{miss}_T >$ 0.2$\times M_{eff}$,
\end{itemize}

\noindent where $S_T$ is the transverse sphericity and $M_{eff}$ is the effective mass.
$M_{eff}$ is caluculated from the leading four jets, $E^{miss}_T$, and the leptons.

\begin{equation}
M_{eff} = \sum^{4}_{i=1}p^{{\rm jet}i}_T + E^{miss}_T (+ \sum_{j}p^{{\rm lepton}j}_T).
\end{equation}

\begin{figure}[h]
\begin{tabular}{cc}
\begin{minipage}{0.5\hsize}
\begin{center} 
\epsfig{file=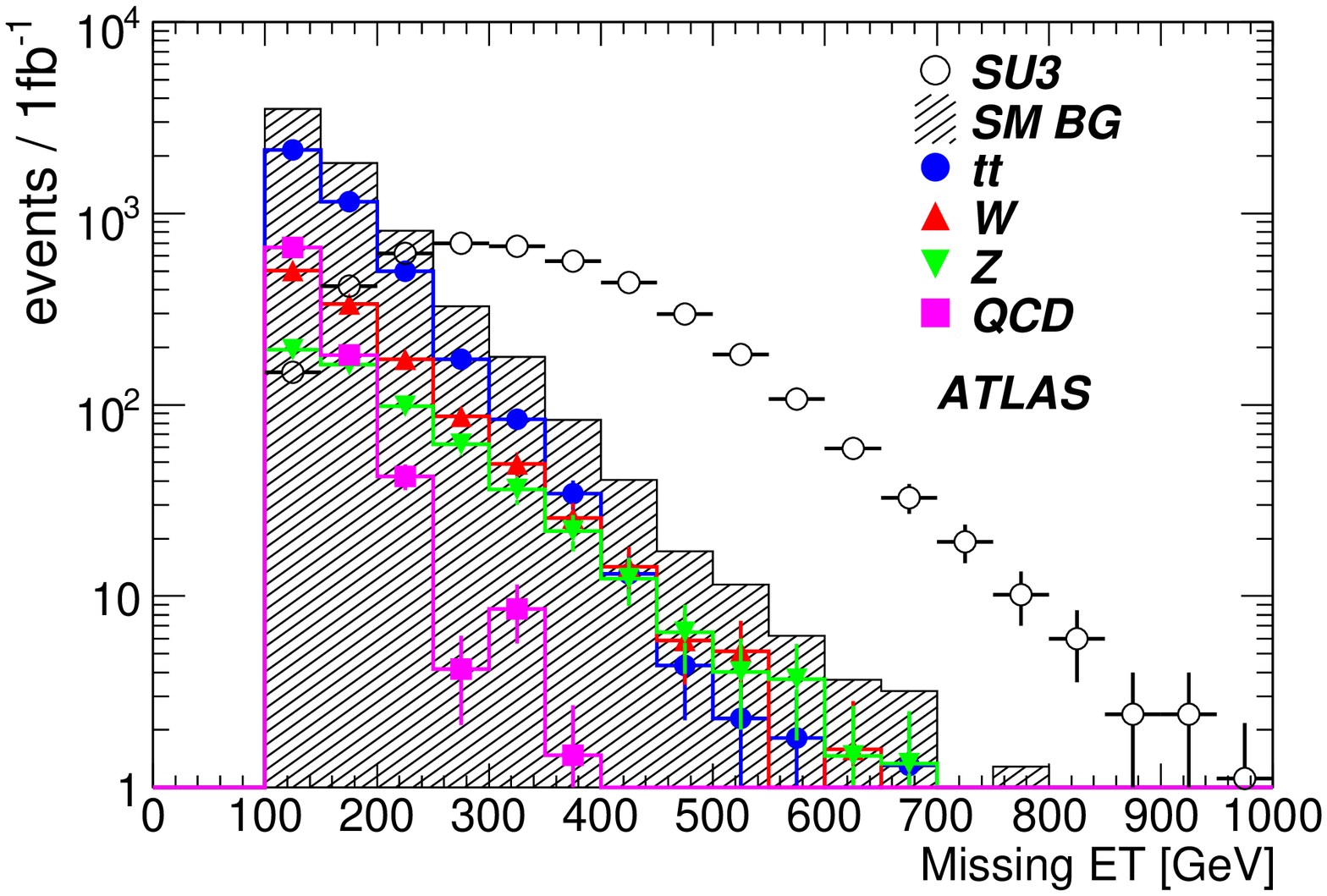, width=0.9\hsize}
\end{center}
\end{minipage}
\begin{minipage}{0.5\hsize}
\begin{center} 
\epsfig{file=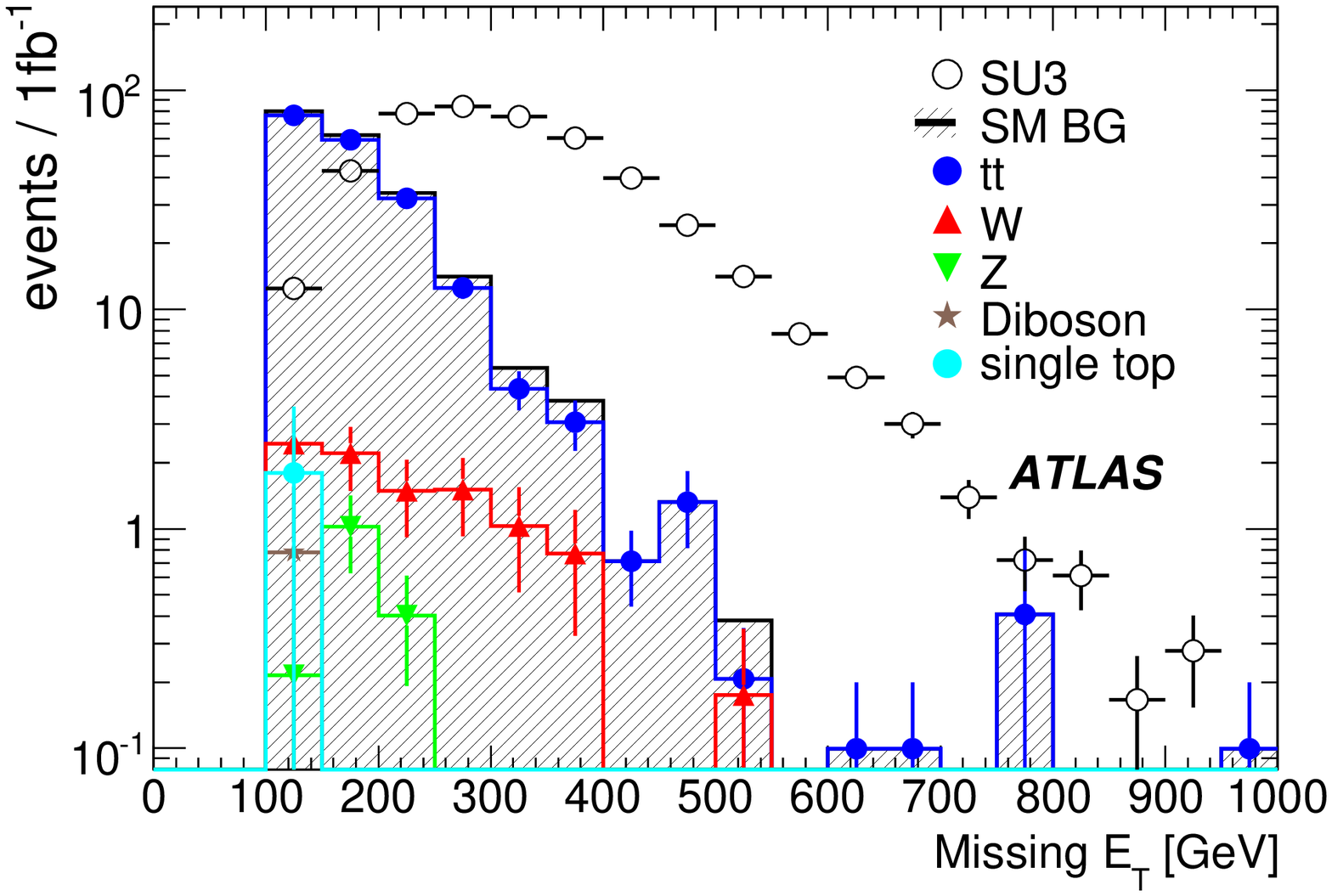, width=0.9\hsize}
\end{center}
\end{minipage}
\end{tabular}
\caption{The $E^{miss}_T$ distributions for no lepton (left) and one lepton (right) mode. The histgrams are normalized to the integrated luminosity of 1fb$^{-1}$.}
\label{fig:mET_withSU3}
\end{figure}

\indent After the selection, the SUSY signal with large $E^{miss}_T$ can be observed as an excess from the SM 
background. Figure \ref{fig:mET_withSU3} shows $E^{miss}_T$ distribution for no lepton search mode. The signal 
shows the promissing SUSY point, where $m_0$ = 100 GeV, $m_{1/2}$ = 300 GeV, 
$A$ = 300 GeV, tan$\beta$ = 6 and $\mu >$ 0. Clear excess over the SM background can be observed with an 
integrated luminosity of 1fb$^{-1}$.

\section{Data-drive Estimation of the Standard Model Backgrounds}
It is essential to understand and estimate the SM background for concrete esvidence of SUSY. We take 
data-driven approach to estimate the SM background because the uncertainties of Monte Carlo predictions 
are expected to be large in the early stage of the experiment.    

\subsection{$M_T$ Method}
\begin{figure}[h]
\begin{tabular}{cc}
\begin{minipage}{0.5\hsize}
\begin{center} 
\epsfig{file=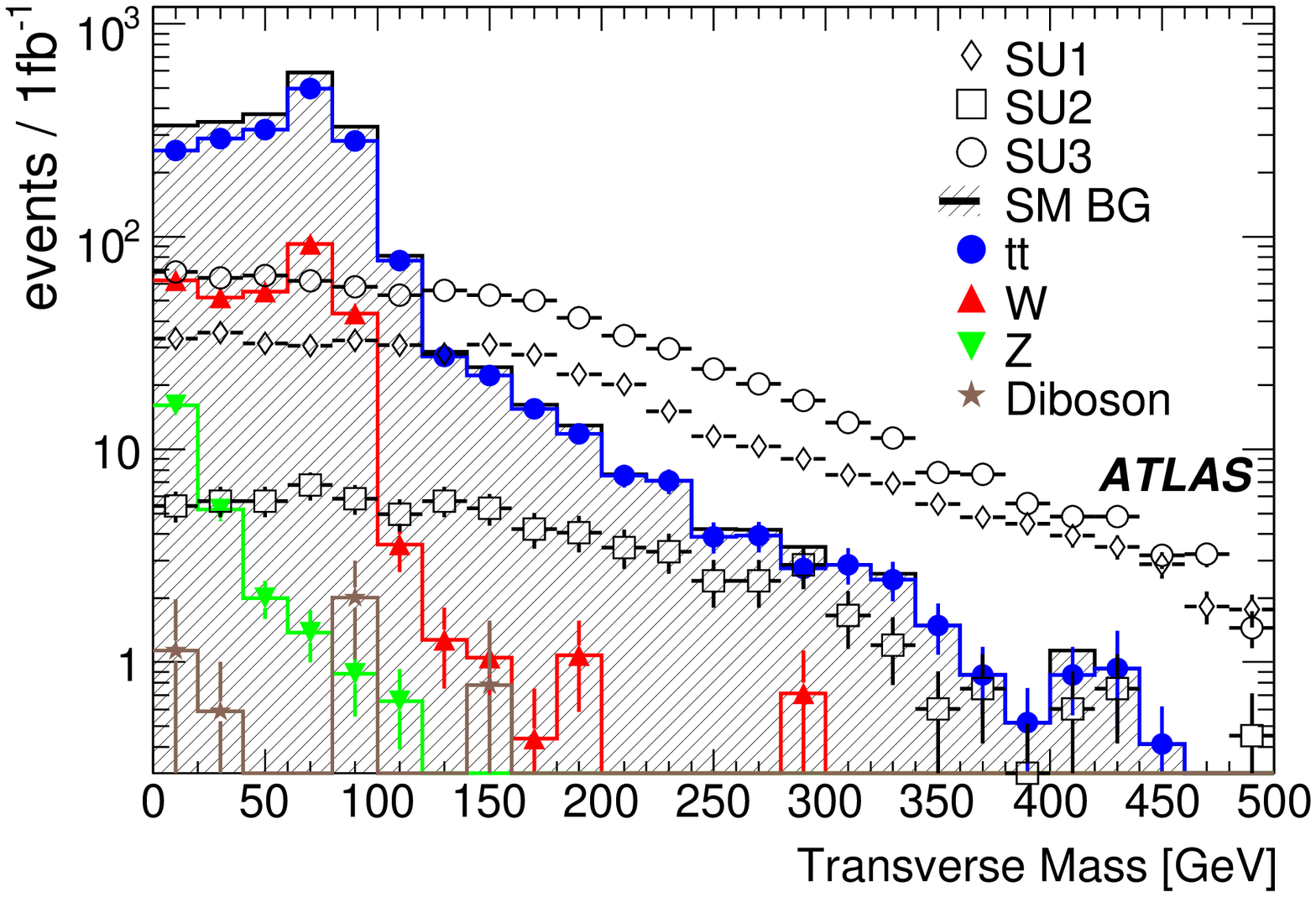, width=0.9\hsize}
\end{center}
\end{minipage}
\begin{minipage}{0.5\hsize}
\begin{center} 
\epsfig{file=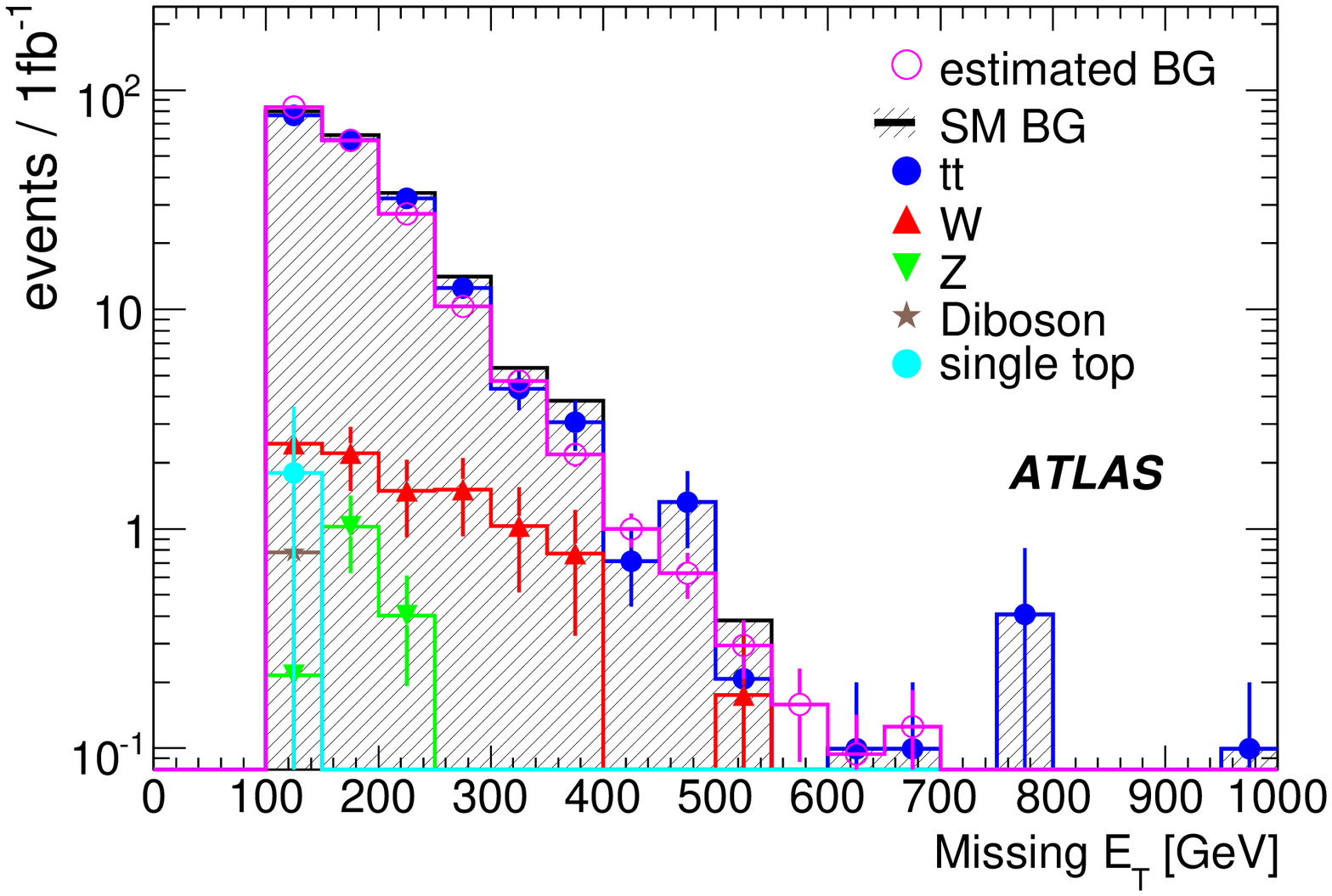, width=0.9\hsize}
\end{center}
\end{minipage}
\end{tabular}
\caption{The $M_T$ distribution (left) and the estimation of $E^{miss}_T$ distribution of the SM background (right) for one lepton SUSY search mode. The histgrams are normalized to the integrated luminosity of 1fb$^{-1}$.}
\label{fig:onelep_estBG}
\end{figure}

Top pair ($t\bar{t}$) and $W^{\pm}$ + jets processes are the dominat backgrounds for one lepton SUSY search mode
(Fig. \ref{fig:mET_withSU3}). For one lepton mode, $M_T >$ 100 GeV is required to
enhance the SUSY signal (Fig. \ref{fig:onelep_estBG}). $M_T$ is the transverse mass between the lepton and
missing energy. The control sample with $M_T < $ 100 GeV is used for the estimation. Since SUSY
contribution in the control sample is much smaller than in the signal region and the shapes of
distributions such as $E^{miss}_T$ are independent of $M_T$, the SM background in the signal region
can be estimated from the control sample as shown in Figure \ref{fig:onelep_estBG}.

\subsection{QCD Background}
\begin{figure}[h]
\begin{tabular}{cc}
\begin{minipage}{0.5\hsize}
\begin{center} 
\epsfig{file=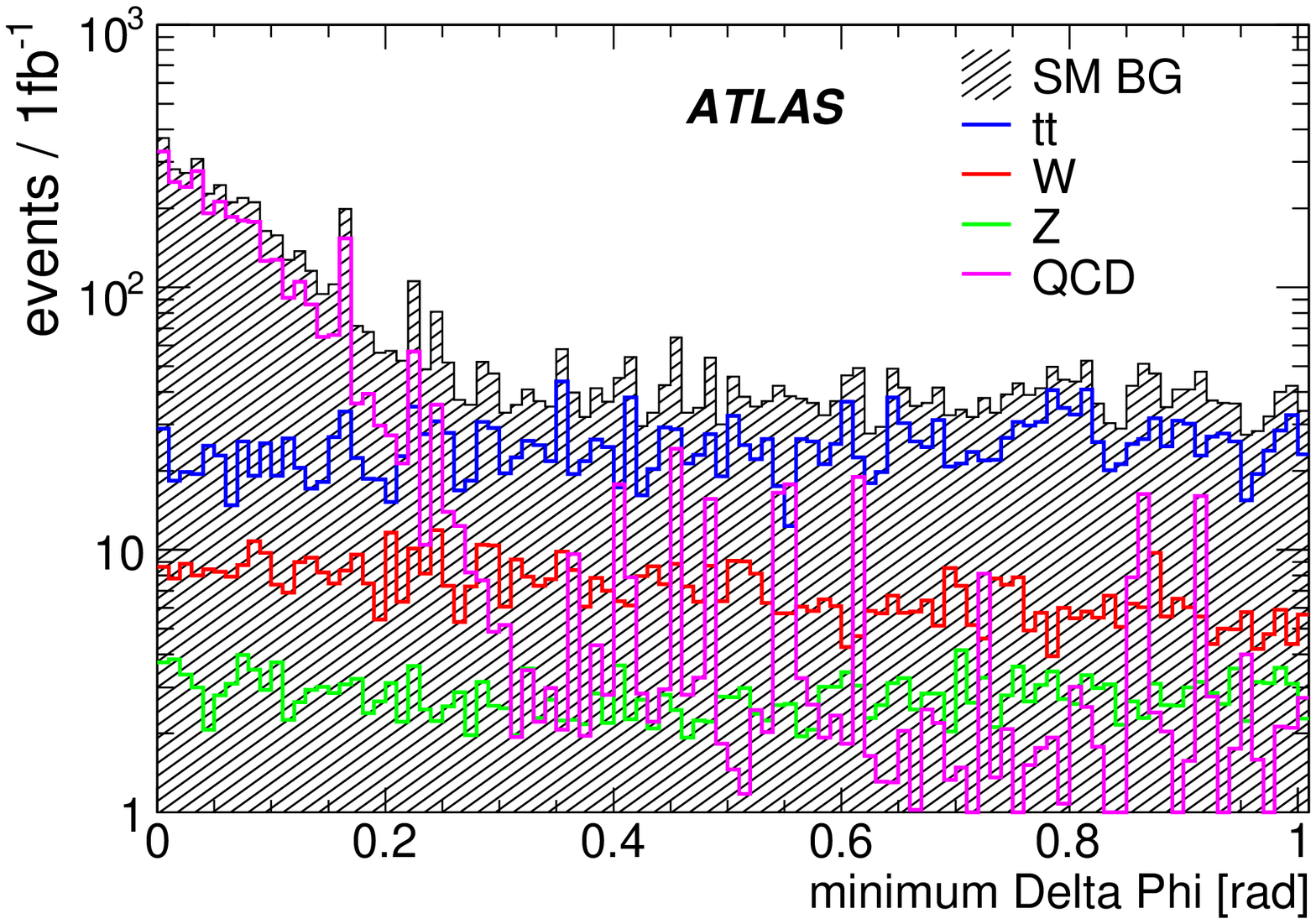, width=0.9\hsize}
\end{center}
\end{minipage}
\begin{minipage}{0.5\hsize}
\begin{center} 
\epsfig{file=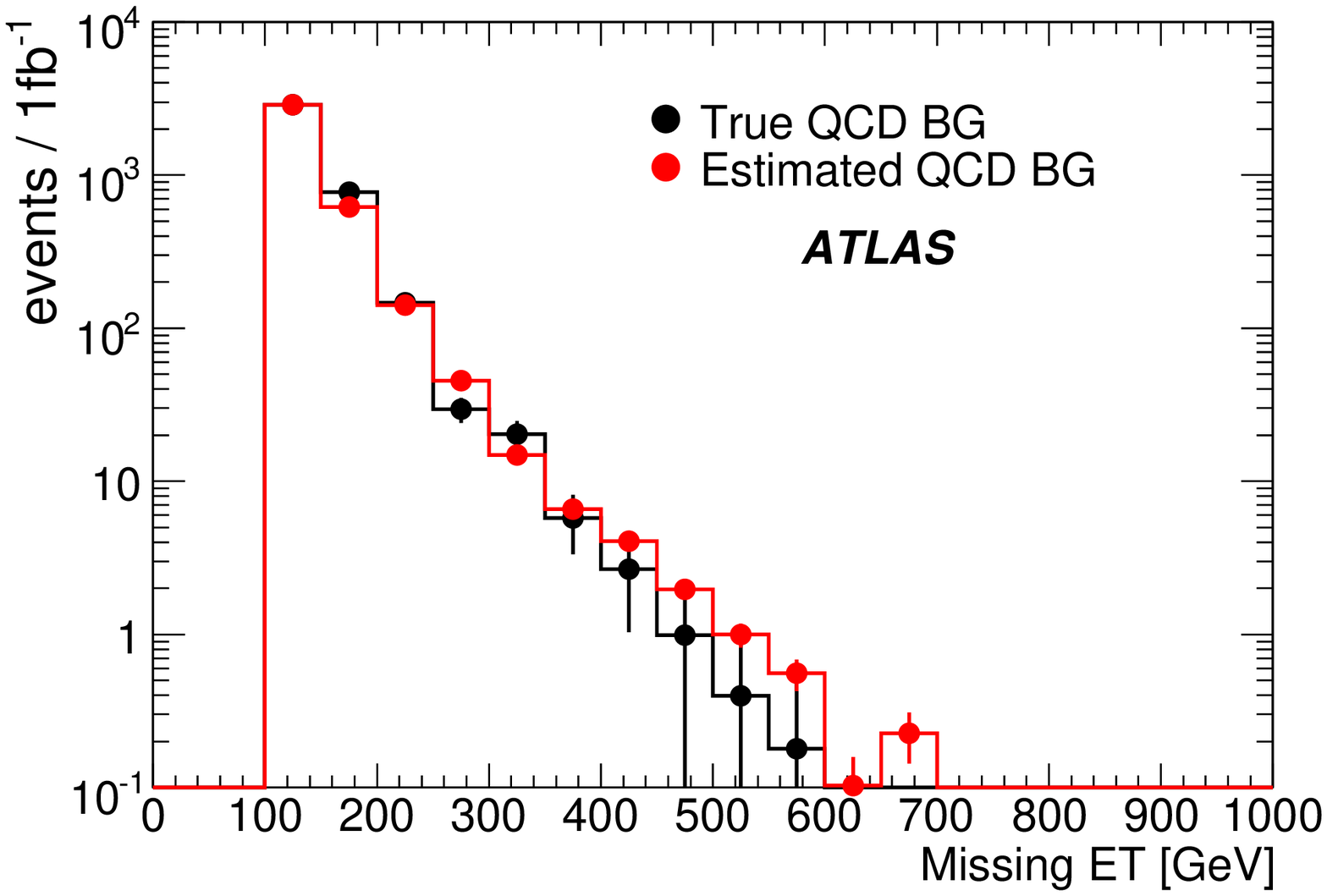, width=0.9\hsize}
\end{center}
\end{minipage}
\end{tabular}
\caption{The minimal $\delta \phi$ distribution (left) and the estimation of $E^{miss}_T$ distribution of the QCD events (right) in no lepton SUSY search mode. The histgrams are normalized to the integrated luminosity of 1fb$^{-1}$.}
\label{fig:nolep_estQCD}
\end{figure}

The QCD processes contribute to the backgrounds of no lepton mode (Fig. \ref{fig:mET_withSU3}). 
In order to reduce the QCD background, we add one more cut which require the 
difference in azimuthal angle $\phi$ between $E^{miss}_T$ and the three most energetic jets to be larger 
than 0.2. The distributions of the minimal value of the $\Delta \phi$ (min $\Delta \phi$) are shown in 
Figure \ref{fig:nolep_estQCD} and the min $\Delta \phi$ distribution of the QCD processes has small 
$\Delta \phi$ because $E^{miss}_T$ in the QCD processes is due to neutrinos 
emitted from semileptonic decays of b/c quarks or mis-measurement of jet energies. \\
\indent The QCD background can be estimated from multi-jet events without 
$E^{miss}_T$. The missing $E_T$ function, which is the momentum fraction of the neutrino in b/c decay, is 
applied to the 1st and 2nd leading jet of the control sample and then the heavy-flavor QCD events with 
large $E^{miss}_T$ can be reproduced. Figure \ref{fig:nolep_estQCD}(b) shows $E^{miss}_T$ distributions of 
true QCD processes and estimation with this method.

\section{Conclusion}
The LHC will start to operate in 2008 and the data-driven method can be used for the background estimation of 
the  SUSY searches in the early stage of the collision. The essence of the data driven approach are summerized 
in this paper. Our searches are based on the SUSY event signitures and less model 
independent but we can discover SUSY whose squarks and gluino masses up to 1 TeV with an integrated 
luminosity of 1fb$^{-1}$.

\section*{Acknowledgments}
The author would like to thank all members of ATLAS and CMS collaborations.

\end{document}